\documentclass[lettersize,journal]{IEEEtran}
\usepackage{amsmath}
\usepackage{threeparttable}
\usepackage{amsmath,amsfonts}
\usepackage{algorithmic}
\usepackage{algorithm}
\usepackage{array}
\usepackage[caption=false,font=normalsize,labelfont=sf,textfont=sf]{subfig}
\usepackage{textcomp}
\usepackage{stfloats}
\usepackage{url}
\usepackage{verbatim}
\usepackage{graphicx}
\usepackage{cite}
\usepackage{setspace}
\usepackage{booktabs}  %  引入三线表宏包
\usepackage{tabularx}
\usepackage{multirow}
\usepackage{cite}
\usepackage{xcolor}  % 已有
\usepackage{hyperref}
\hypersetup{
    colorlinks=true, 
    linkcolor=blue!60!black,
    citecolor=blue!60!black,
    urlcolor=blue!60!black,  
    filecolor=blue!60!black
}
\usepackage{bm}
\usepackage[nameinlink]{cleveref}
\usepackage{float}%提供float浮动环境
\begin{document}
\bibliographystyle{IEEEtran}

% \title{Enabling Fair Allocation in Power Systems with Scalable Shapley: The Case of Carbon Emission Responsibility in Power Networks}
\title{Deep Learning-Accelerated Shapley Value for Fair Allocation in Power Systems: The Case of Carbon Emission Responsibility}

\author{
Yuanhao Feng,~\IEEEmembership{Student Member,~IEEE}, Tao Sun, Yan Meng, Xuxin Yang, Donghan Feng,~\IEEEmembership{Senior Member,~IEEE}

\thanks{This work was sponsored in part by the Smart Grid-National Science and Technology Major Project (2024ZD0800400), in part by the Science and Technology Project of State Grid Corporation of China (52094024004P), and in part by the Science and Technology Commission of Shanghai Municipality (23XD1422000). 
%\textit{(Yuanhao Feng and
%Tao Sun contributed equally to this work.)  %(Corresponding authors: Tao Sun; Donghan %Feng.)}}% <-this % stops a space
\textit{(Yuanhao Feng and
Tao Sun contributed equally to this work.)}}% <-this % stops a space

\thanks{Yuanhao Feng, Yan Meng, Xuxin Yang, and Donghan Feng are with the School of Electrical Engineering, Shanghai Jiao Tong University, Shanghai 200240, China (e-mail: fyh386884223@sjtu.edu.cn; meng\_yan@sjtu.edu.cn; yangxuxin@sjtu.edu.cn; seed@sjtu.edu.cn).
}

\thanks{Tao Sun is with the School of Electrical Engineering, Shanghai Jiao Tong University, Shanghai 200240, China, and also with the Department of Civil and Environmental Engineering, Stanford University, California, 94305 USA (e-mail: luke18@stanford.edu).
}
}


\IEEEpubid{}

\maketitle

\begin{abstract}

Allocating costs, benefits, and emissions fairly among power system participant entities represents a persistent challenge. The Shapley value provides an axiomatically fair solution, yet computational barriers have limited its adoption beyond small-scale applications. This paper presents SurroShap, a scalable Shapley value approximation framework combining efficient coalition sampling with deep learning surrogate models that accelerate characteristic function evaluations. 
Exemplified through carbon emission responsibility allocation in power networks, SurroShap enables Shapley-based fair allocation for power systems with thousands of entities for the first time. We derive theoretical error bounds proving that time-averaged SurroShap allocations converge to be $\varepsilon$-close to exact Shapley values. Experiments on nine systems ranging from 26 to 1,951 entities demonstrate completion within the real-time operational window even at maximum scale, achieving $10^4-10^5$× speedups over other sampling-based methods while maintaining tight error bounds. The resulting Shapley-based carbon allocations possess six desirable properties aligning individual interests with decarbonization goals. Year-long simulations on the Texas 2000-bus system validate real-world applicability, with regional analysis revealing how renewable-rich areas offset emission responsibility through exports while load centers bear responsibility for driving system-wide generation.

\end{abstract}


\section{Introduction}
\IEEEPARstart{F}{air} allocation of costs, benefits, emissions, and resources among diverse entities in power systems has been a long-standing challenge \cite{CREMERS2023120328}. As the power industry undergoes transformative changes such as energy transition, market deregulation, and distributed generation proliferation, the number of participating entities has expanded dramatically \cite{local_electricity_markets}. This includes not only traditional generators and utilities but also prosumers, aggregators, virtual power plants, and energy communities \cite{COSTA2025115597,9811386}, making fair allocation among these diverse and numerous entities increasingly complex yet critical for system operation and market efficiency.

The Shapley value represents the game-theoretic gold standard for fair allocation, determining each entity's contribution by averaging their marginal impact across all possible coalitions \cite{shapley1953value}. This solution concept has demonstrated its versatility across diverse application domains, ranging from infrastructure and business operations such as telecommunications cost-sharing \cite{KIM201655}, risk capital allocation \cite{HOMBURG2008208}, and supply chain profit distribution \cite{kemahliouglu2011centralizing}, to its recent emergence as a fundamental tool for feature importance quantification in explainable artificial intelligence \cite{kumar2020problems}. The Shapley value's reputation for fairness stems from its unique position as the only allocation method simultaneously satisfying four fundamental axioms: efficiency, symmetry, additivity, and dummy player \cite{young1985monotonic}. However, implementing Shapley values for complex real-world systems faces a critical computational barrier: calculating exact values requires evaluating all $2^n$ possible coalitions, making the method intractable beyond modest entity counts.

To address this exponential complexity, researchers have developed various approximation strategies. Early work introduced Monte Carlo sampling of random permutations to estimate marginal contributions \cite{mann1960values}. Subsequent advances improved convergence through stratified \cite{castro2017improving} and quasi-Monte Carlo \cite{mitchell2022sampling} methods, alongside tailored allocation strategies including Neyman allocation \cite{mathew2013efficiency} and empirical Bernstein sampling \cite{burgess2021approximating}. The recent surge in explainable AI has catalyzed development of sophisticated approximation methods such as Kernel SHAP \cite{lundberg2017unified}, Leverage SHAP\cite{musco2025provably}, and Gradient SHAP \cite{erion2020learning}, enabling efficient Shapley value computation for high-dimensional machine learning models.

In power systems, Shapley value and its computationally efficient approximations have found diverse applications including congestion cost allocation \cite{VOSWINKEL2022119039}, network loss distribution \cite{SHARMA201733,sharma2016loss}, generation start-up cost sharing \cite{1626369}, energy community benefit distribution \cite{8226810,CREMERS2023120328,valencia2025stable}, microgrid cost allocation \cite{9619937}, demand response and virtual power plant profit distribution \cite{o2015shapley,WANG2021114180,DABBAGH2015368}, transmission expansion cost and benefit allocation \cite{9601291,kristiansen2018mechanism}, and carbon emission responsibility (CER) assignment \cite{zhou2019cooperative}. However, these Shapley-based allocations have been predominantly limited to small-scale systems: most implementations involve fewer than 20 entities \cite{1626369,8226810,valencia2025stable,9619937,WANG2021114180,DABBAGH2015368,9601291,kristiansen2018mechanism,zhou2019cooperative,VOSWINKEL2022119039,o2015shapley}, with only a few reaching several dozen entities \cite{SHARMA201733,sharma2016loss}. A rare exception is \cite{CREMERS2023120328}, which handles 200 entities but relies on application-specific consumer clustering to reduce the actual number of entities being allocated. This scalability limitation stems from a unique computational challenge in power system allocations beyond coalition explosion: determining each coalition's characteristic function often requires solving optimization problems, equilibrium computations, or systems of equations \cite{maleki_addressing_2015}. For instance, CER allocation requires solving an optimal power flow (OPF) for each coalition to determine generation dispatch and resulting emissions; generation start-up cost sharing necessitates security-constrained unit commitment optimization; and transmission planning cost allocation demands solving models that minimize operational and investment costs. While solving a single such problem is computationally trivial, Shapley value computation for larger systems can require evaluating tens or even hundreds of millions of coalitions, creating a compound computational barrier that has restricted practical implementations to small-scale systems even when efficient sampling techniques are employed.

We propose SurroShap, a novel approximation method for Shapley allocations in power systems that synergistically combines KernelSHAP's efficient coalition sampling \cite{covert2021improving} with deep neural network surrogate models that rapidly approximate characteristic function evaluations, addressing both computational barriers simultaneously. We demonstrate this approach through CER allocation among generators and loads in power networks, where the characteristic function (total system carbon emissions) normally requires solving an OPF. The proposed framework readily extends to other power system allocation problems by training appropriate surrogate models for their respective characteristic functions.

This paper makes two primary contributions. First, we enable accurate Shapley value-based fair allocation for large-scale power systems, achieving speedups of tens of thousands of times compared to existing sampling-based methods and completing allocation for thousands of entities within the 5-minute real-time operational window. Second, we advance fair CER allocation, which is critical for climate change mitigation, by providing fairness benchmarks for existing methods (carbon emission flow \cite{kang2015carbon,7741580}, marginal carbon intensity \cite{valenzuela2023dynamic}, Aumann-Shapley \cite{xie2024real,chen2018method}) and scaling Shapley-based CER allocation from dozens to thousands of entities, enabling deployment in real-world power systems.

The remainder of this paper is organized as follows. Section \ref{Section2} formulates the fair CER allocation problem using Shapley values. Section \ref{Section3} presents the SurroShap methodology, including its theoretical error bounds. Section \ref{Section4} provides comprehensive case studies across multiple systems. Section \ref{Section5} concludes the paper and discusses future research directions.

\section{Fair allocation of carbon emission responsibility}\label{Section2}
\subsection{Problem Formulation}\label{Section2A}

This paper studies the problem of allocating the responsibility of carbon emissions from power generation among all generators and loads in the power network in a fair manner based on the Shapley value\cite{shapley1953value}, the well-recognized fair allocation method from cooperative game theory. The premise of such allocation is that carbon emissions are not merely the responsibility of generators themselves but also of loads, as electricity consumption is the ultimate driver of generation and thus emissions \cite{zhou2019cooperative}.

The fair allocation is formulated as follows. The total carbon emissions for a given time period (typically an hour to align with power system operations) is determined by the generation output (MWh) of each thermal generator multiplied by its carbon emission intensity ($\mathrm{tCO_2eq/MWh}$). The generation dispatch is determined through the OPF, which maximizes social welfare while respecting system constraints.

To allocate emissions using the Shapley value, we consider all possible coalitions of system entities. A coalition $S$ represents a subset of generators and loads participating in the system, while the network infrastructure remains unchanged. For each coalition, we solve a new OPF to determine the resulting generation dispatch and corresponding emissions. The Shapley value for each entity is then calculated as its average marginal contribution across all possible coalitions it could join, weighted by the probability of each coalition forming.

Specifically, we consider three types of entities: thermal generation units ($G$), renewable generation units ($R$), and loads ($D$). For a coalition $S\subseteq N$, where $N = G \cup R \cup D$ is the set of all entities, the DC OPF problem is formulated as:
\begin{align}
\label{1}\tag{1}
\min\limits_{P_{gt}^{\rm G},\,P_{rt}^{\rm R},\,P_{dt}^{\rm D}}
&\;\;\sum_{g\in G\cap S}\rho_{gt}^{\rm G} P_{gt}^{\rm G}
 \;-\; \sum_{d\in D\cap S}\rho^{\rm VOLL} P_{dt}^{\rm D}
\end{align}

\noindent
s.t.
\allowdisplaybreaks
\begin{align}
\label{2}\tag{2}
&\sum_{g\in G\cap S} P_{gt}^{\rm G}
 + \sum_{r\in R\cap S} P_{rt}^{\rm R}
= \sum_{d\in D\cap S} P_{dt}^{\rm D} \\[6pt]
\label{3}\tag{3}
& P_{g}^{\rm Gmin} \;\leq\; P_{gt}^{\rm G} \;\leq\; P_{g}^{\rm Gmax},
&& \forall g \in G\cap S \\[6pt]
\label{4}\tag{4}
& 0 \;\leq\; P_{rt}^{\rm R} \;\leq\; P_{rt}^{\rm Rmax},
&& \forall r \in R\cap S \\[6pt]
\label{5}\tag{5}
& 0 \;\leq\; P_{dt}^{\rm D} \;\leq\; P_{dt}^{\rm Dmax},
&& \forall d \in D\cap S \\[6pt]
& -U_f \;\leq\;
   \sum_{g\in G\cap S} F_{fg}^{\rm G} P_{gt}^{\rm G}
 + \sum_{r\in R\cap S} F_{fr}^{\rm R} P_{rt}^{\rm R} \notag \\[-2pt]
\label{6}\tag{6}
& \quad\quad
 - \sum_{d\in D\cap S} F_{fd}^{\rm D} P_{dt}^{\rm D}
 \;\leq\; U_f,
&& \forall f
\end{align}
where $P_{gt}^{\rm G}$, $P_{rt}^{\rm R}$, and $P_{dt}^{\rm D}$ are the decision variables representing the power output of thermal unit $g$, renewable unit $r$, and power demand served for load $d$ at time $t$, respectively. The objective function \eqref{1} maximizes social welfare, where $\rho_{gt}^{\rm G}$ is the offer price of thermal unit $g$ at time $t$, and $\rho^{\rm VOLL}$ is the value of lost load. Note that this formulation implicitly assumes renewable units have zero offer prices, reflecting their zero marginal cost nature. Additionally, we assume $\rho^{\rm VOLL}$ to be sufficiently large,  treating loads as inelastic in this paper. Constraint \eqref{2} ensures power balance. Constraints \eqref{3}-\eqref{5} enforce the operational limits, where $P_{g}^{\rm Gmin}$ and $P_{g}^{\rm Gmax}$ are the minimum and maximum power outputs of thermal unit $g$, $P_{rt}^{\rm Rmax}$ is the maximum available power from renewable unit $r$ at time $t$, and $P_{dt}^{\rm Dmax}$ is the maximum power demand of load $d$ at time $t$. Constraint \eqref{6} represents the transmission security constraints, where $F_{fg}^{\rm G}$, $F_{fr}^{\rm R}$, and $F_{fd}^{\rm D}$ are the power transfer distribution factors from generators and loads to branch $f$, and $U_{f}$ is the capacity limit of branch $f$.

After solving the OPF for coalition $S$, we calculate the characteristic function value, which is the total carbon emissions of coalition $S$ at time $t$:
\begin{align}
	\label{7}\tag{7}&c_t(S)=\sum_{g\in G\cap S}\beta_{g}P_{gt}^{\rm G}
\end{align}
where $\beta_{g}$ is the carbon emission intensity of thermal unit $g$ ($\mathrm{tCO_2eq/MWh}$).

The Shapley value for entity $i$ at time $t$, representing its allocated CER, is then computed as:
\begin{align}
\label{8}\tag{8}&x_{it} = \sum_{S \subseteq N \setminus i} \frac{n_S! \, (n_N - n_S - 1)!}{n_N!} \left[ c_t(S \cup \{i\}) - c_t(S) \right]
\end{align}
where $n_{S}$ is the number of entities in coalition $S$  and $n_{N}$ is the total number of entities. The term $c_t(S \cup \{i\}) - c_t(S)$ represents entity $i$'s marginal contribution to emissions when joining coalition $S$. The weighting factor $\frac{n_S! \, (n_N - n_S - 1)!}{n_N!}$ accounts for all possible orderings in which coalitions can form. This formulation ensures that each entity's allocation reflects its average impact on system emissions across all possible participation scenarios.

\subsection{Allocation Properties}\label{Section2B}
The Shapley-based CER allocation, in addition to inheriting the well-known fairness guarantee, exhibits several meaningful properties that align individual incentives with system-wide emission reduction objectives. These properties emerge naturally from the cooperative game framework and demonstrate the allocation's effectiveness as an instrument for carbon mitigation. We summarize six key properties below, with empirical validation provided in Section \ref{Section4} and theoretical proofs under mild assumptions presented in Appendix A in the supplementary file.

\begin{enumerate}
\item \textbf{Property 1 (Clean Energy Incentive):} Renewable generation units receive non-positive CERs, reflecting their contribution to system decarbonization.

\item \textbf{Property 2 (Emission Intensity Improvement):} Thermal generators that reduce their carbon emission intensity (efficiency improvements, carbon capture technologies, etc.) receive lower responsibility allocations.

\item \textbf{Property 3 (Merit Order Alignment):} Low-emission thermal generators that reduce their offer prices experience decreased CERs.

\item \textbf{Property 4 (Load Responsibility):} Electricity loads receive non-negative CERs, acknowledging that consumption drives generation and associated emissions.

\item \textbf{Property 5 (Conservation Incentive):} Load reduction directly translates to decreased CER allocation.

\item \textbf{Property 6 (Temporal Flexibility Value):} For entities with uncertain or variable output (renewables and loads), profile reshaping that maintains total energy could simultaneously reduce both individual carbon responsibilities and system-wide emissions.
\end{enumerate}

\section{The SurroShap Method}\label{Section3}

Although the Shapley value provides theoretically fair allocation of CER, two computational barriers prevent its practical deployment in power systems\cite{maleki_addressing_2015}. First, the number of coalitions grows exponentially with the number of entities—specifically, $2^n$ coalitions for $n$ entities. Second, computing the characteristic function $c_t(S)$ for each coalition compounds the computational burden; in CER allocation, this requires solving an OPF problem for each coalition. We propose SurroShap, an efficient approximation method that enables near real-time Shapley value computation even for systems with thousands of entities on standard computing resources. SurroShap addresses the coalition explosion problem through KernelSHAP\cite{lundberg2017unified}, a sampling-based approximation method, and accelerates characteristic function evaluation using a deep neural network surrogate model. We derive theoretical bounds on the approximation error between SurroShap and exact Shapley values (computed via exhaustive enumeration of all coalitions), and demonstrate that SurroShap is $\varepsilon$-close to exact Shapley values under specific conditions.

\subsection{KernelSHAP: Sampling-Based Shapley Approximation}\label{Section3A}
Sampling-based methods have been widely adopted to accelerate Shapley value computation in various domains. KernelSHAP, a weighted least-squares approach, demonstrates superior convergence to exact Shapley values as the number of sampled coalitions increase, compared with traditional Monte Carlo sampling, and has been widely used for Shapley-based feature importance in explainable AI \cite{lundberg2017unified,covert2021improving}. 

KernelSHAP approximates Shapley values by solving the following weighted regression problem:
\begin{align}
	\label{9}\tag{9}&\hat{\mathbf{x}}_{t} = \arg\min_{\hat{x}_{it}} \ \mathbb{E}_{S\sim p(S)}\left[\left( c_t(S) - \sum_{i\in S}\hat{x}_{it} \right)^2 \right]
\end{align}
subject to:
\begin{align}
	\label{10}\tag{10}&\sum_{i\in N}\hat{x}_{it}=c_t(N)
\end{align}
where $\hat{x}_{it}$ denotes the KernelSHAP estimate for entity $i$ at time $t$, $\hat{\mathbf{x}}_{t} = [\hat{x}_{1t}, \hat{x}_{2t}, \dots, \hat{x}_{n_N t}]^\top$
 is the vector of all estimates, and the sampling distribution $p(S)$ is defined as:
\begin{align}
	\label{11}\tag{11}&p(S) \propto \frac{n_N - 1}{\binom{n_N}{n_S} n_S  (n_N - n_S)}
\end{align}

After sampling $M$ coalitions, the optimization problem becomes:
\begin{align}
    \label{12}\tag{12}&\hat{\mathbf{x}}_{t} = \arg\min_{\hat{x}_{it}} \frac{1}{M} \sum_{m=1}^{M}\left( c_t(S_m) - \sum_{i\in S_m}\hat{x}_{it} \right)^2
\end{align}
where $S_m$ denotes the $m$-th sampled coalition. The state-of-the-art implementation employs paired sampling \cite{covert2021improving}, where $\mathbf{s}_{M/2+m}=\mathbf{e}-\mathbf{s}_m$ for $m \leq M/2$, with $\mathbf{s}_m$ being the binary indicator vector for coalition $S_m$ (i.e., $[\mathbf{s}_m]_i = 1$ if $i \in S_m$, else 0) and $\mathbf{e}$ being the vector of ones. This yields the analytical solution:
\begin{align}
    \label{13}\tag{13}&\hat{\mathbf{x}}_t = \hat{\mathbf{A}}_t^{-1} \left[ \hat{\mathbf{b}}_t - \frac{\mathbf{e}^T \hat{\mathbf{A}}_t^{-1} \hat{\mathbf{b}}_t - c_t(N)}{\mathbf{e}^T \hat{\mathbf{A}}_t^{-1} \mathbf{e}}\mathbf{e} \right]
\end{align}
where:
\begin{align}
    \label{14}\tag{14}&\hat{\mathbf{A}}_{t} = \frac{1}{M}\sum_{m=1}^{M}\mathbf{s}_m\mathbf{s}_m^T, \quad \hat{\mathbf{b}}_{t} = \frac{1}{M}\sum_{m=1}^{M}c_t(S_m)\mathbf{s}_m
\end{align}

\subsection{Deep Learning Surrogate Model}\label{Section3B}

The primary computational bottleneck in KernelSHAP lies in evaluating $c_t(S_m)$ for all sampled coalitions, which requires solving the OPF problem \eqref{1}-\eqref{6} in our case. With hundreds of millions of coalition samples needed for convergence as the entity count grows, this becomes computationally prohibitive. To accelerate this process, we employ a deep neural network (DNN) as a surrogate model that directly maps coalition configurations and system operating conditions to carbon emissions without explicitly solving the OPF.

We first generate a comprehensive dataset by solving the OPF problem under diverse operating conditions. Specifically, we vary thermal unit offer prices, renewable maximum outputs, load maximum demands, and coalition configurations while maintaining fixed network topology and transmission constraints. Each OPF solution yields the corresponding carbon emission, creating input-output pairs for supervised learning.

We train a multi-layer feed-forward DNN following the architecture in \cite{9205647}, comprising linear layers with ReLU activations. The network takes as input the thermal unit offer prices $\bm{\rho}_{t}^{\rm G}$, carbon intensities $\bm{\beta}$, renewable maximum outputs $\mathbf{P}_{t}^{\rm Rmax}$, load maximum demands $\mathbf{P}_{t}^{\rm Dmax}$, and coalition indicator vector $\mathbf{s}_{m}$, where bold notation denotes the column vectors of respective parameters. The output is the estimated carbon emission $c^*_t(S_m)$ for coalition $S_m$.

% \begin{figure}[!t]
% \centering
% \includegraphics[width=88mm]{Figures/Surromodel.pdf}
% \caption{Schematic of deep learning surrogate model. A DNN is employed as a surrogate for model \eqref{2}-\eqref{8}, with the parameters $\bm{\rho}_{t}^{\rm G},\bm{\beta},\mathbf{P}_{t}^{\rm Rmax},\mathbf{P}_{t}^{\rm Dmax},\mathbf{s}_{m}$ that may vary as inputs, and $c_t(S_m)$ as the output.}
% \label{Surromodel}
% \end{figure}

The trained surrogate model is expressed as:
\begin{align}
    \label{15}\tag{15}&c^*_t(S_m)=\mathcal{F}_\theta\left(\bm{\rho}_{t}^{\rm G},\bm{\beta},\mathbf{P}_{t}^{\rm Rmax},\mathbf{P}_{t}^{\rm Dmax},\mathbf{s}_{m}\right)
\end{align}
where $\mathcal{F}_\theta(\cdot)$ is the function realized by the DNN with parameters $\theta$.

\subsection{SurroShap Algorithm}\label{Section3C}
\begin{figure}[!t]
\centering
\includegraphics[width=88mm]{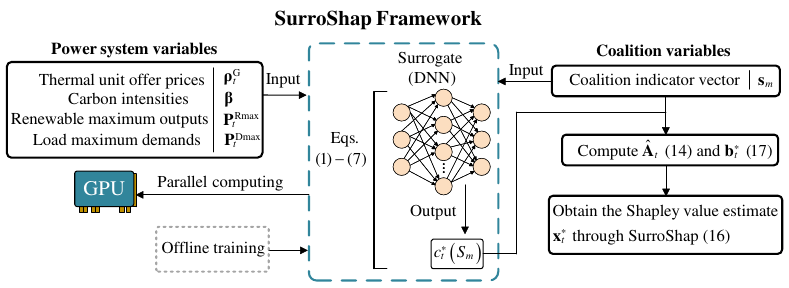}
\caption{Schematic diagram of the SurroShap framework.}
\label{fig:surroshap_framework}
\end{figure}

SurroShap integrates KernelSHAP with the DNN surrogate model by replacing the computationally expensive OPF evaluations with rapid neural network inference. Fig. \ref{fig:surroshap_framework} illustrates the overall framework, demonstrating how power system operating conditions and coalition samples are processed through the DNN surrogate to generate Shapley-based CER allocations.

The complete computational procedure is presented in Algorithm \ref{alg:surroshap}. Similar to KernelSHAP's formulation in \eqref{13}, SurroShap computes the allocation as:
\begin{align}
    \label{16}\tag{16}&\mathbf{x}^*_t = \hat{\mathbf{A}}_t^{-1} \left[ \mathbf{b}^*_t - \frac{\mathbf{e}^T \hat{\mathbf{A}}_t^{-1} \mathbf{b}^*_t - c_t(N)}{\mathbf{e}^T \hat{\mathbf{A}}_t^{-1} \mathbf{e}}\mathbf{e} \right]
\end{align}
where $\mathbf{b}^*_t$ is analogous to $\hat{\mathbf{b}}_t$ but computed using DNN-estimated emissions:
\begin{align}
    \label{17}\tag{17}&\mathbf{b}^*_t = \frac{1}{M}\sum_{m=1}^{M}c^*_t(S_m)\mathbf{s}_m
\end{align}

The algorithm's efficiency stems from two key features: 1) the large number of DNN inferences in steps 8–12 can be accelerated through GPU parallel computing, and 2) KernelSHAP's efficient sampling strategy provides accurate approximations to exact Shapley values with substantially fewer samples needed than Monte Carlo sampling methods.

\begin{algorithm}
\caption{SurroShap Computational Procedure}
\label{alg:surroshap}
\begin{algorithmic}[1]
\STATE \textbf{Input:} Trained DNN with parameters $\theta$; system parameters $\bm{\rho}_{t}^{\rm G}$, $\bm{\beta}$, $\mathbf{P}_{t}^{\rm Rmax}$, $\mathbf{P}_{t}^{\rm Dmax}$ for all time steps $t \in \{1, \ldots, T\}$
\STATE \textbf{Output:} SurroShap allocations $\mathbf{x}^*_t$ for all time steps
\STATE \textbf{Initialize:} $t \leftarrow 1$
\WHILE{$t \leq T$}
    \STATE Initialize $\hat{\mathbf{A}}_t \leftarrow \mathbf{0}$, $\mathbf{b}^*_t \leftarrow \mathbf{0}$
    \STATE Sample first $M/2$ coalition vectors $\{\mathbf{s}_1, \ldots, \mathbf{s}_{M/2}\}$ from distribution $p(S)$ in \eqref{11}
    \STATE Generate paired samples: $\mathbf{s}_{M/2+m} \leftarrow \mathbf{e} - \mathbf{s}_m$ for $m = 1, \ldots, M/2$
    \FOR{$m = 1$ to $M$}
        \STATE Compute $c^*_t(S_m) \leftarrow \mathcal{F}_\theta(\bm{\rho}_{t}^{\rm G}, \bm{\beta}, \mathbf{P}_{t}^{\rm Rmax}, \mathbf{P}_{t}^{\rm Dmax}, \mathbf{s}_{m})$
        \STATE Update $\hat{\mathbf{A}}_t \leftarrow \hat{\mathbf{A}}_t + \frac{1}{M}\mathbf{s}_m\mathbf{s}_m^T$
        \STATE Update $\mathbf{b}^*_t \leftarrow \mathbf{b}^*_t + \frac{1}{M}c^*_t(S_m)\mathbf{s}_m$
    \ENDFOR
    \STATE Compute $\mathbf{x}^*_t$ using \eqref{16} with $\hat{\mathbf{A}}_t$ and $\mathbf{b}^*_t$
    \STATE $t \leftarrow t + 1$
\ENDWHILE
\RETURN $\{\mathbf{x}^*_1, \ldots, \mathbf{x}^*_{T}\}$
\end{algorithmic}
\end{algorithm}

\subsection{Approximation Error Analysis}\label{Section3D}
While SurroShap can achieve dramatic computational acceleration, it introduces approximation errors from the DNN surrogate model that prevent the convergence guarantee inherent in KernelSHAP. In this subsection, we derive theoretical bounds on SurroShap's approximation error and demonstrate that under regularity conditions, the time-averaged multi-period allocations, which align with standard power system operational practice, converge to be $\varepsilon$-close to exact Shapley values.

\subsubsection{Single-Period Error Bound}
For a single time period $t$, we decompose the total approximation error between SurroShap and exact Shapley values using the triangle inequality:
\begin{equation}
\label{18}\tag{18}
\|\mathbf{x}_{t}-\mathbf{x}^*_t\| \leq \|\mathbf{x}_{t}-\hat{\mathbf{x}}_{t}\| + \|\hat{\mathbf{x}}_{t}-\mathbf{x}^*_t\| \leq \eta_t + \|\hat{\mathbf{x}}_{t}-\mathbf{x}^*_t\|
\end{equation}
where $\|\cdot\|$ denotes the L2 norm, $\eta_t$ represents KernelSHAP's inherent approximation error bound (which converges to zero as sample size $M$ increases), and $\|\hat{\mathbf{x}}_{t}-\mathbf{x}^*_t\|$ captures the additional error introduced by the DNN surrogate.

To characterize the surrogate-induced error, we define the conditional bias vector $\bm{\delta}_{t}$ with elements:
\begin{equation}
\label{19}\tag{19}
[\bm{\delta}_{t}]_i = \mathbb{E}_m[c^*_t(S_m)-c_t(S_m)|i\in S_m]
\end{equation}
representing the expected prediction bias when entity $i$ participates in sampled coalitions. From  \eqref{13} and \eqref{16}, the difference between KernelSHAP and SurroShap estimates arises solely from their respective $\mathbf{b}$ vectors:
\begin{equation}
\label{20}\tag{20}
\hat{\mathbf{b}}_{t}-\mathbf{b}^*_t = \frac{1}{M}\sum_{m=1}^{M}\left[c_t(S_m)-c^*_t(S_m)\right]\mathbf{s}_m = \frac{1}{2}\bm{\delta}_{t}
\end{equation}
Substituting this relationship yields:
\begin{equation}
\label{21}\tag{21}
\|\hat{\mathbf{x}}_{t}-\mathbf{x}^*_t\| = \left\|\frac{1}{2}\left(\hat{\mathbf{A}}_t^{-1}-\frac{\hat{\mathbf{A}}_t^{-1}\mathbf{e}\mathbf{e}^T \hat{\mathbf{A}}_t^{-1}}{\mathbf{e}^T \hat{\mathbf{A}}_t^{-1} \mathbf{e}}\right)\bm{\delta}_{t}\right\|
\end{equation}

\subsubsection{Multi-Period Error Bound}
In practical power system operations, allocations are typically aggregated across multiple time periods (e.g., hourly allocations summed for daily or monthly billing). The time-averaged error bound over $T$ periods becomes:
\begin{equation}
\label{22}\tag{22}
\left\|\sum_{t=1}^{T}\frac{\mathbf{x}_{t}-\mathbf{x}^*_t}{T}\right\| 
\leq 
\left\|\sum_{t=1}^{T}\frac{\mathbf{x}_{t}-\hat{\mathbf{x}}_{t}}{T}\right\| 
+ 
\left\|\sum_{t=1}^{T}\frac{\hat{\mathbf{x}}_{t}-\mathbf{x}^*_t}{T}\right\|
\end{equation}
As the number of samples $M$ increases, the matrix $\hat{\mathbf{A}}_t$ converges to a constant matrix $\tilde{\mathbf{A}}$ independent of the specific coalition samples\cite{covert2021improving}, with elements given by:
\begin{equation}
\label{23}\tag{23}
[\tilde{\mathbf{A}}]_{ii} = \tfrac{1}{2}, \ \forall i
\qquad
[\tilde{\mathbf{A}}]_{ij} =
\frac{
  \sum\nolimits^{n_N-1}_{z=2} \frac{z-1}{n_N - z}
}{
  \sum\nolimits^{n_N-1}_{z=1} \frac{n_N (n_N - 1)}{z (n_N - z)}
}, \ \forall i \neq j
\end{equation}
This convergence property allows us to establish an asymptotic bound. Define the conditional mean bias error (MBE) vector across all time periods as:
\begin{equation}
\label{24}\tag{24}
\bm{\epsilon}=\mathbb{E}_{t}\left[\bm{\delta}_{t}\right]=\frac{1}{T}\sum_{t=1}^{T}\bm{\delta}_{t}
\end{equation}
For large $M$ and $T$, this represents the DNN surrogate's systematic bias pattern. The asymptotic error bound then becomes:
\begin{equation}
\label{25}\tag{25}
\left\|\sum_{t=1}^{T}\frac{\mathbf{x}_{t}-\mathbf{x}^*_t}{T}\right\| \leq \eta + \varepsilon
\end{equation}
where $\eta = \sum_{t=1}^{T}\eta_t/T$ is the average KernelSHAP error bound (which vanishes as $M$ increases), and:
\begin{equation}
\label{26}\tag{26}
\varepsilon=\left\|\frac{1}{2}\left(\tilde{\mathbf{A}}^{-1}-\frac{\tilde{\mathbf{A}}^{-1}\mathbf{e}\mathbf{e}^T \tilde{\mathbf{A}}^{-1}}{\mathbf{e}^T \tilde{\mathbf{A}}^{-1} \mathbf{e}}\right)\bm{\epsilon}\right\|
\end{equation}

\subsubsection{Practical Error Estimation}
The bound $\varepsilon$ depends directly on the DNN surrogate's conditional MBE. We make the assumption that $|\text{MBE}| \ll \text{RMSE}$, which is reasonable since $|\text{MBE}| \leq \text{RMSE}$ by definition and basic algebra. The assumption is then validated empirically in Section \ref{Section4}. Under this assumption, a properly trained surrogate model ensures small $\varepsilon$. This establishes that time-averaged SurroShap allocations are asymptotically $\varepsilon$-close to exact Shapley values. It is important to note that for a single time period, the conditional bias vector $\bm{\delta}_t$ exhibits temporal variability and cannot be reliably inferred from the DNN's MBE. This is why we establish the $\varepsilon$-closeness guarantee only for multi-period averages, where the aggregation over time periods allows $\bm{\epsilon}$ to have a stable pattern that can be characterized by the DNN's overall performance metrics.

To estimate the average KernelSHAP error
bound $\eta$ in practice without computing intractable exact Shapley values, we exploit KernelSHAP's convergence behavior. For the error bound component $\eta_t$, we fit a power-law decay function with logarithmic correction to the convergence trajectory:
\begin{equation}
\label{27}\tag{27}
\begin{aligned}
\phi_t(m) &= \|\hat{\mathbf{x}}_{t}^{(M-\Delta M +m)}-\hat{\mathbf{x}}_{t}^{(M-\Delta M)}\| \\
&= \lambda_t + \frac{\kappa_t}{m^{\alpha_t}\ln(m+\gamma_t)}
\end{aligned}
\end{equation}
where $\hat{\mathbf{x}}_{t}^{(k)}$ denotes the KernelSHAP estimate using $k$ samples, $\Delta M$ defines the tail interval for fitting, and $\lambda_t, \kappa_t, \alpha_t, \gamma_t$ are fitted parameters. Since $\phi_t(m)\to  \|\mathbf{x}_{t}-\hat{\mathbf{x}}_{t}^{(M-\Delta M)}\| \geq \|\mathbf{x}_{t}-\hat{\mathbf{x}}_{t}\|$ as $m \to \infty$, this provides a practical estimate of the KernelSHAP error bound component $\eta_t$.

The complete error bound can thus be estimated by combining $\eta$ with the conditional MBE $\bm{\epsilon}$ computed from the DNN's test set performance, enabling practical verification that SurroShap achieves the required accuracy for operational deployment.

\section{Numerical Experiments}\label{Section4}
This section presents comprehensive numerical experiments to validate SurroShap's performance across multiple dimensions: computational efficiency, approximation accuracy, and allocation properties. We conduct experiments on nine test systems ranging from small academic benchmarks of around 26 entities to real-world scale applications of nearly 2000 entities. All optimization problems for training data generation are solved on a high-performance computing (HPC) cluster, while SurroShap inference and allocation computations are performed on a single workstation equipped with two GPUs. The software implementation uses Python 3.9.0 with Gurobi 10.0.1 for optimization and PyTorch 1.11.0 for deep learning.

\subsection{Test Systems and Data Sources}\label{Section4A}
We evaluate SurroShap on nine power systems with increasing number of entities involved (see Table \ref{Computational efficiency}): IEEE 30-bus, IEEE 39-bus, IEEE 57-bus, IEEE 24-bus, Central Illinois 200-bus, IEEE 118-bus, IEEE 300-bus, South Carolina 500-bus, and Texas 2000-bus systems. System parameters including generator limits, transmission line reactances, and capacities are sourced from the electric grid test cases published by Texas A\&M University \cite{tamu_electric_grid_test_cases}. Carbon emission intensities are set at 1.044 tCO$_2$eq/MWh for coal units and 0.44 tCO$_2$eq/MWh for gas units, derived from EIA data \cite{eia_faq_74} by dividing total emissions by generation.

A separate DNN surrogate model is trained for each test system following the methodology in Section \ref{Section3B}. All nine systems are used to demonstrate SurroShap's computational efficiency and scalability for single-period allocation. For deeper analysis, we select three representative systems spanning different scales:

\begin{itemize}
\item \textbf{IEEE 30-bus}: Small-scale validation where exact Shapley values remain tractable, enabling direct accuracy verification and comparison with alternative sampling-based approximation methods.
\item \textbf{IEEE 118-bus}: Medium-scale demonstration of visual analysis of CER distribution and validation of allocation properties.
\item \textbf{Texas 2000-bus}: Large-scale real-world application with year-long simulation demonstrating practical deployment feasibility.
\end{itemize}

For multi-period analysis, operating conditions are generated as follows. In the IEEE 30-bus system, a two-week (336-hour) simulation uses hourly thermal offers varied by $\pm20\%$ uniform random factors from cost coefficients, renewable maximum outputs scaled between 0 and 1.2 times representative profiles, and load demands between 0.4 and 1.2 times base values, with all base quantities sourced from \cite{tamu_electric_grid_test_cases}. The IEEE 118-bus system employs seasonal representative days with renewable capacities and locations from NREL data \cite{7904729}, using 2023 ERCOT wind/solar profile shapes \cite{ercot_pg7_126_m} scaled to IEEE 118-bus installed capacities, and 2023 load profile shapes from ERCOT archives \cite{ercot_hourly_load_archives} scaled to match system load levels \cite{tamu_electric_grid_test_cases}. The Texas 2000-bus system uses year-long data with thermal offers adjusted by EIA coal/gas price variations \cite{eia_coal_markets,eia_natgas_weekly_update}, actual ERCOT renewable profiles \cite{ercot_pg7_126_m}, and weather-region-specific load data \cite{ercot_hourly_load_archives} calibrated to the test system scenarios \cite{tamu_electric_grid_test_cases} and year of 2023.

\subsection{Deep Learning Surrogate Training}\label{Section4B}
We train DNN surrogate models that map coalition configurations and system operating conditions to carbon emissions as detailed in Section \ref{Section3B} for each of the test systems. A separate DNN surrogate model is trained for each test system. To generate the dataset for training, validating, and testing each DNN model, we create 10 million data samples. Each sample consists of input variables as specified in \eqref{15} along with the corresponding carbon emission output obtained by solving the OPF with these inputs. The input variables are sampled from realistic distributions calibrated based on the system parameters specified in Section \ref{Section4A}, with thermal offer prices, renewable maximum outputs, load maximum demands, and carbon intensities varied within feasible operational ranges, while the coalition indicator vector $\mathbf{s}_{m}$ is randomly sampled from all possible coalitions. These randomly sampled input variables are combined to form each data sample, with this process repeated to generate all 10 million samples. The dataset is then randomly split into 70\% training (7 million), 20\% validation (2 million), and 10\% test (1 million) partitions.

Training employs the Adam optimizer with an initial learning rate of $5 \times 10^{-4}$ that decreases by factor 0.3 every 5 epochs over 50 total epochs, weight decay of $1 \times 10^{-4}$, and mean squared error (MSE) loss function. We conduct hyperparameter tuning, which results in an architecture consisting of 8 fully connected layers with 512 neurons each and ReLU activations. The trained DNN models' performance on their respective test sets for the three representative systems is presented in Table \ref{Computational error}, showing root mean squared error (RMSE), MBE, and R-squared values.

\subsection{Computational Performance}\label{Section4C}
Table \ref{Computational efficiency} also demonstrates SurroShap's dramatic computational advantages across systems of varying scale by summarizing single-period (hourly) allocations. While exact Shapley computation becomes intractable beyond around 30 entities (requiring over $10^5$ minutes for 31 entities), SurroShap maintains sub-minute computation even for the South Carolina with 290 entities and completes the 1,951-entity Texas system allocation in 3.17 minutes, meeting real-time operational requirements. Notably, SurroShap's computation time exhibits non-monotonic behavior between 157-290 entities (0.32 to 0.26 minutes), as I/O and other fixed overhead dominate computations at this scale. The number of sampled coalitions we used for different system scales increases with the number of entities, selected by the criterion that 10\% additional samples change the L2 norm of estimated Shapley values by less than 0.1\%. KernelSHAP, while significantly improving upon exact Shapley, still requires over two months ($10^5$ minutes) for the Texas system. SurroShap achieves a $10^4$-$10^5$ speedup over KernelSHAP through GPU-accelerated DNN inference, enabling practical deployment at scale.

\begin{table*}[!t]
\caption{Computational Performance of SurroShap Across Different System Scales}
\centering
\small
\setlength{\tabcolsep}{2.5pt}
\begin{threeparttable}
\begin{tabular}{|l|c|c|c|c|c|c|c|c|c|}
\toprule
\textbf{System} & \textbf{IEEE} & \textbf{IEEE} & \textbf{IEEE} & \textbf{IEEE} & \textbf{Central Illinois} & \textbf{IEEE} & \textbf{IEEE} & \textbf{South Carolina} & \textbf{Texas} \\
& \textbf{30-bus} & \textbf{39-bus} & \textbf{57-bus} & \textbf{24-bus} & \textbf{200-bus} & \textbf{118-bus} & \textbf{300-bus} & \textbf{500-bus} & \textbf{2000-bus} \\
\midrule
Number of entities $n_N$ & 26 & 31 & 49 & 50 & 157 & 171 & 260 & 290 & 1,951 \\
exact Shapley (minutes) & $3,032$ & $10^5$ & $(10^{10})$ & $(10^{11})$ & $(10^{43})$ & $(10^{47})$ & $(10^{74})$ & $(10^{83})$ & $(10^{580})$ \\
KernelSHAP (minutes) & $723$ & $799$ & $857$ & $1,416$ & $2,836$ & $4,114$ & $4,511$ & $7,403$ & $10^5$ \\
SurroShap (minutes) & $0.08$ & $0.09$ & $0.11$ & $0.16$ & $0.32$ & $0.33$ & $0.26$ & $0.26$ & $3.17$ \\
Sampled coalitions (million) & $4$ & $4$ & $4$ & $7$ & $7$ & $10$ & $10$ & $10$ & $100$ \\
\bottomrule
\end{tabular}
\begin{tablenotes}
\footnotesize
\item[]Note: Values in parentheses indicate estimates. Times reported for single-machine computation with 2 GPUs and 16 CPU cores.
\end{tablenotes}
\end{threeparttable}
\label{Computational efficiency}
\end{table*}

\subsection{Error Bounds and Convergence Analysis}\label{Section4D}
\begin{figure}[!t]
\centering
\includegraphics[width=88mm]{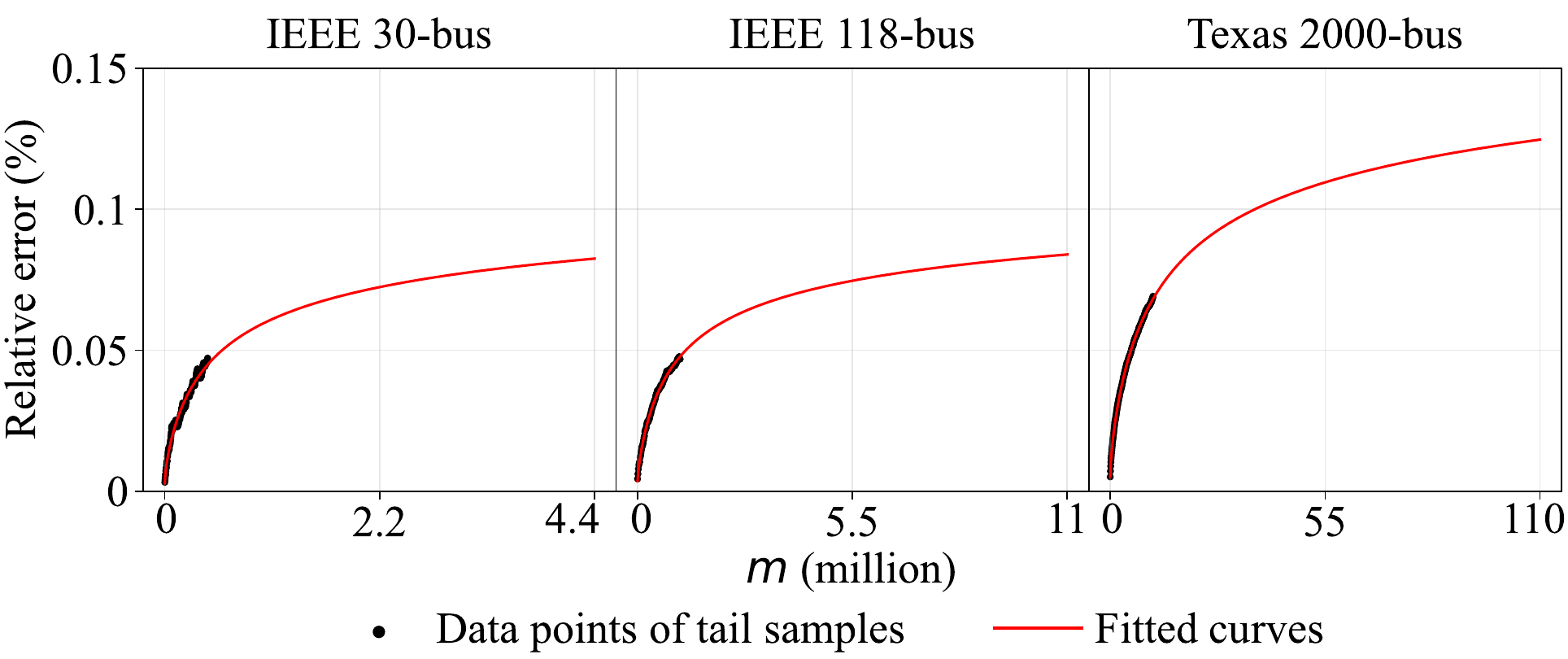}
\caption{Estimation of KernelSHAP approximation error bounds via power-law decay function fitting to convergence trajectories. See $\phi_t(m)$ in \eqref{27}.}
\label{fit_kernel_error}
\end{figure}

We establish SurroShap's approximation accuracy through theoretical bounds calculated based on Section \ref{Section3D}, which includes two components: the KernelSHAP approximation error and the DNN surrogate-induced error. To estimate the KernelSHAP error bound component $\eta_t$ for single-period (hourly) allocation, we fit convergence trajectories using the power-law decay function defined in equation \eqref{27}. This fitting process is performed for three representative systems by using the tail 10\% of sampled coalitions to extract asymptotic error bounds. As shown in Fig. \ref{fit_kernel_error}, the fitted function converges to specific values as $m \to \infty$, providing system-specific KernelSHAP error bound components (relative to the L2 norm of  KernelSHAP values) of 0.18\% for IEEE 30-bus, 0.17\% for IEEE 118-bus, and 0.28\% for Texas 2000-bus systems in this single period.

Table \ref{Computational error} summarizes error statistics for both the DNN surrogate and SurroShap across the three representative systems. The increasing RMSE of the DNN for larger-scale systems is expected as they generally have higher absolute emission values; however, the R-squared values indicate consistent relative accuracy across all scales. The MBEs are orders of magnitude smaller than RMSEs, validating the assumption made in Section \ref{Section3D} that $|\text{MBE}| \ll \text{RMSE}$. The multi-period theoretical error bounds shown in Table \ref{Computational error} are calculated based on \eqref{25} and computed relative to the L2 norm of estimated Shapley values, guaranteeing cumulative multi-period errors below 0.37\%, 0.41\%, and 2.22\% for the three systems respectively.

\begin{table}[!t]
\caption{SurroShap Error Analysis}
\centering
\small
\setlength{\tabcolsep}{3pt}
\begin{tabular}{lccc}
\toprule
\raisebox{0.5\height}{\textbf{Metric}}
             & \shortstack{\textbf{IEEE} \\ \textbf{30-bus}} 
             & \shortstack{\textbf{IEEE} \\ \textbf{118-bus}} 
             & \shortstack{\textbf{Texas} \\ \textbf{2000-bus}} \\
\midrule
DNN RMSE & 1.61 & 43.17 & 321.91 \\
DNN MBE & 0.0009 & 0.0664 & 0.3502 \\
DNN R-squared & 0.9995 & 0.9964 & 0.9980 \\
Theoretical bound (\%) & 0.3688 & 0.4119 & 2.2193 \\
\bottomrule
\end{tabular}
\label{Computational error}
\end{table}

\begin{figure}[!t]
\centering
\includegraphics[width=88mm]{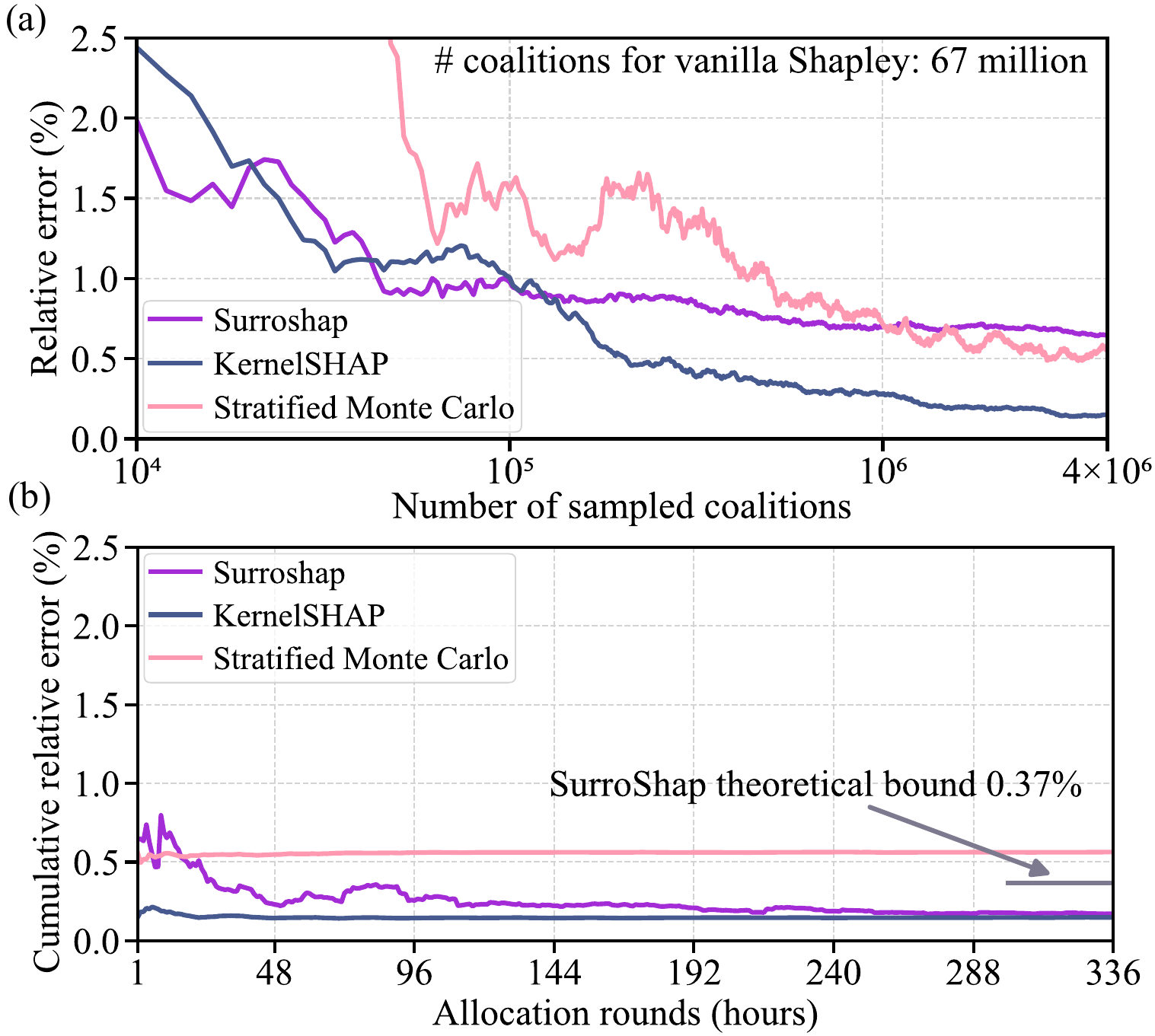}
\caption{Convergence of approximation errors on IEEE 30-bus system. (a) Single-period relative error as a function of number of sampled coalitions. (b) Cumulative error evolution across multiple allocation rounds.}
\label{methed_fig_combine}
\end{figure}

Fig. \ref{methed_fig_combine} demonstrates the convergence properties of SurroShap's approximation error compared to its theoretical bound and shows how errors evolve with increasing coalition sample size and across multiple allocation rounds. Since calculating actual empirical error requires exact Shapley values, we present results only for the IEEE 30-bus system. The comparison includes SurroShap, KernelSHAP\cite{covert2021improving}, and stratified Monte Carlo sampling methods\cite{castro2017improving}.

For single-period (hourly) allocation, as the number of sampled coalitions increases to 4 million, SurroShap converges to 0.64\% error while KernelSHAP achieves 0.15\% and stratified Monte Carlo 0.56\%. The multi-period aggregation, computed with 4 million sampled coalitions, demonstrates the theoretical property established in Section \ref{Section3D}: after two weeks (336 hourly allocation rounds), the cumulative error of SurroShap reduces to 0.17\%, well below the theoretical bound of 0.37\%, while KernelSHAP and stratified Monte Carlo remain at 0.15\% and 0.56\%. The relative errors are computed as the L2 norm of the difference between estimated Shapley values and exact Shapley values, normalized by the L2 norm of the estimated Shapley values.

\subsection{Validation of Allocation Properties}\label{Section4E}
\begin{figure*}[!t]
\centering
\includegraphics[width=180mm]{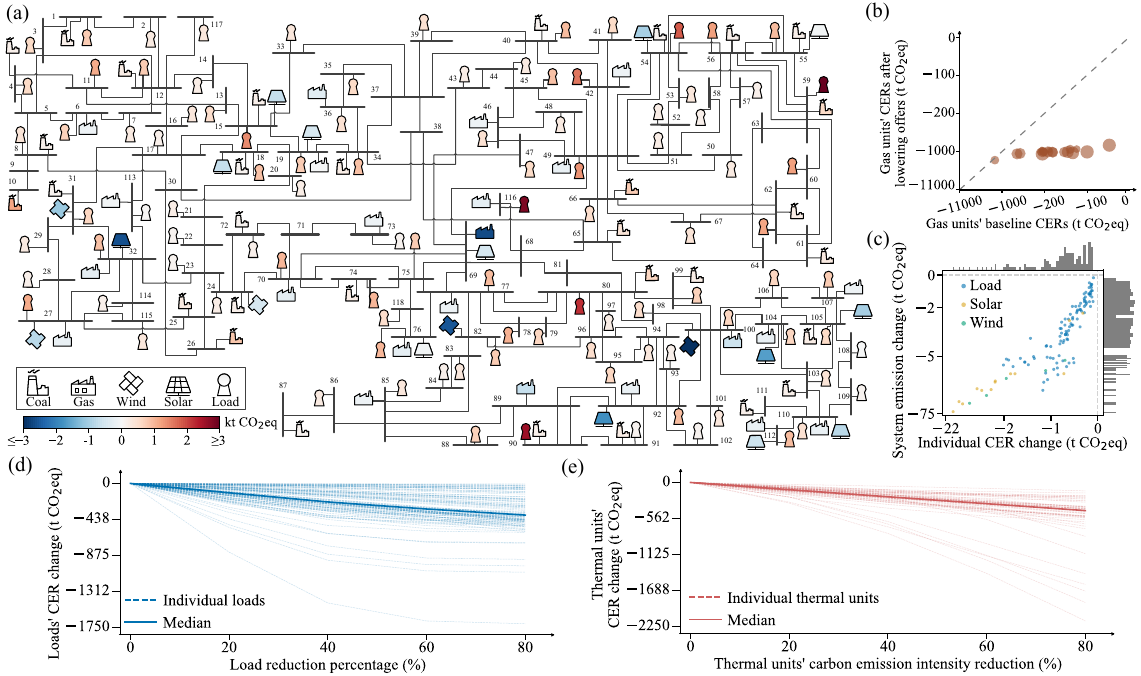}
\caption{Validation of allocation properties on IEEE 118-bus system. (a) Baseline CER distribution. (b) Property 3: reduced gas offers decrease CERs. (c) Property 6: profile reshaping benefits. (d) Property 5: load reduction incentive. (e) Property 2: emission intensity improvement rewards.}
\label{combine118}
\end{figure*}

The IEEE 118-bus system serves to empirically validate the six properties established in Section \ref{Section2B}. Fig. \ref{combine118}(a) presents baseline CER allocation results obtained from daily allocations (sum from hourly allocations) averaged across four representative days (one per season), confirming Properties 1 and 4: renewable units receive negative CERs while loads incur positive CERs.

Fig. \ref{combine118}(b) validates Property 3 by individually reducing each gas unit's offer price and computing the resulting CER allocation. Each point represents one gas unit, plotting its baseline CER against its CER after offer reduction. All points fall below the identity line, confirming that improved market competitiveness of lower-emission thermal units reduces their allocated CERs.

Fig. \ref{combine118}(c) demonstrates Property 6 by reshaping renewable and load profiles. For each entity, we heuristically search for alternative hourly profiles that maintain total energy but adjust individual hours by up to ±2\% of capacity (renewables) or peak demand (loads). The figure shows that for all renewable units and loads, we identify reshaped profiles that simultaneously reduce both individual CERs and total system emissions.

Sensitivity analyses in Figs. \ref{combine118}(d)-(e) confirm Properties 2 and 5 where we examine one entity each time while others remain unchanged. Proportional reductions in load demand or thermal emission intensity consistently decrease the corresponding entity's CER, with median trends (solid lines) and individual responses (dashed) showing robust monotonic relationships consistent with the stated properties.

\subsection{Annual Analysis of Large-Scale System}\label{Section4F}
\begin{figure}[!t]
\centering
\includegraphics[width=88mm]{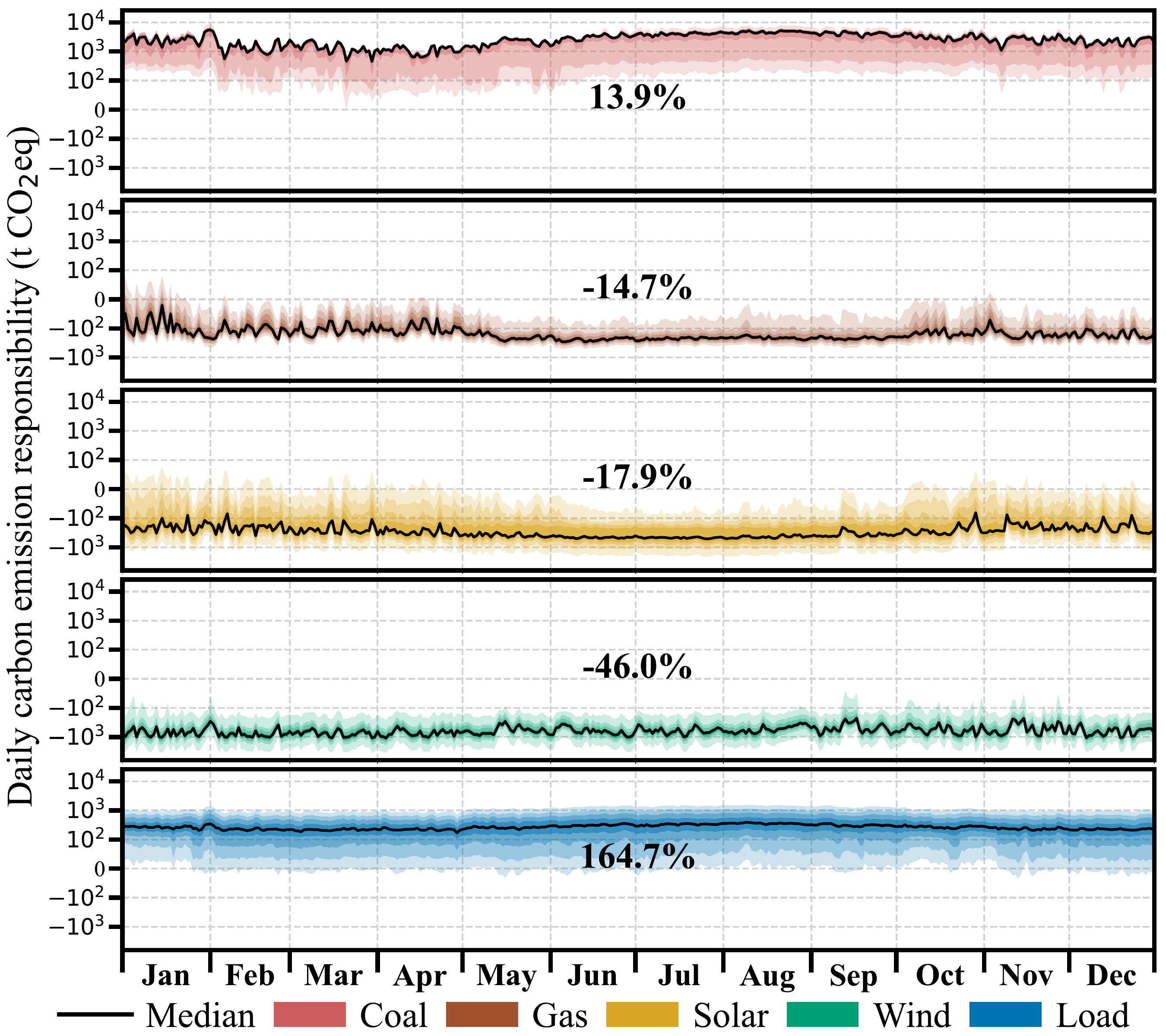}
\caption{Year-long CER allocation for Texas 2000-bus system showing daily statistics by entity type. Median values (solid lines) with percentile ranges (10th-90th through 40th-60th) demonstrate seasonal patterns and allocation stability.}
\label{2000_allagents_fanchart}
\end{figure}

The Texas 2000-bus system demonstrates SurroShap's real-world applicability through year-long simulation of 1,951 entities over 8,760 hourly periods. Fig. \ref{2000_allagents_fanchart} presents daily CER allocations (aggregated from hourly results) for each entity type over a year. The figure displays the median daily CER (solid line) and percentile ranges (10th-90th percentile as lightest shading through 40th-60th percentile as darkest shading) throughout the year. Aggregating all entities' annual CERs and computing each type's percentage of total system emissions reveals: loads account for 164.7\% of total emissions, renewable generation (solar and wind) provides -63.9\%, while coal and gas units contribute 13.9\% and -14.7\% respectively.

\begin{figure}[!t]
\centering
\includegraphics[width=88mm]{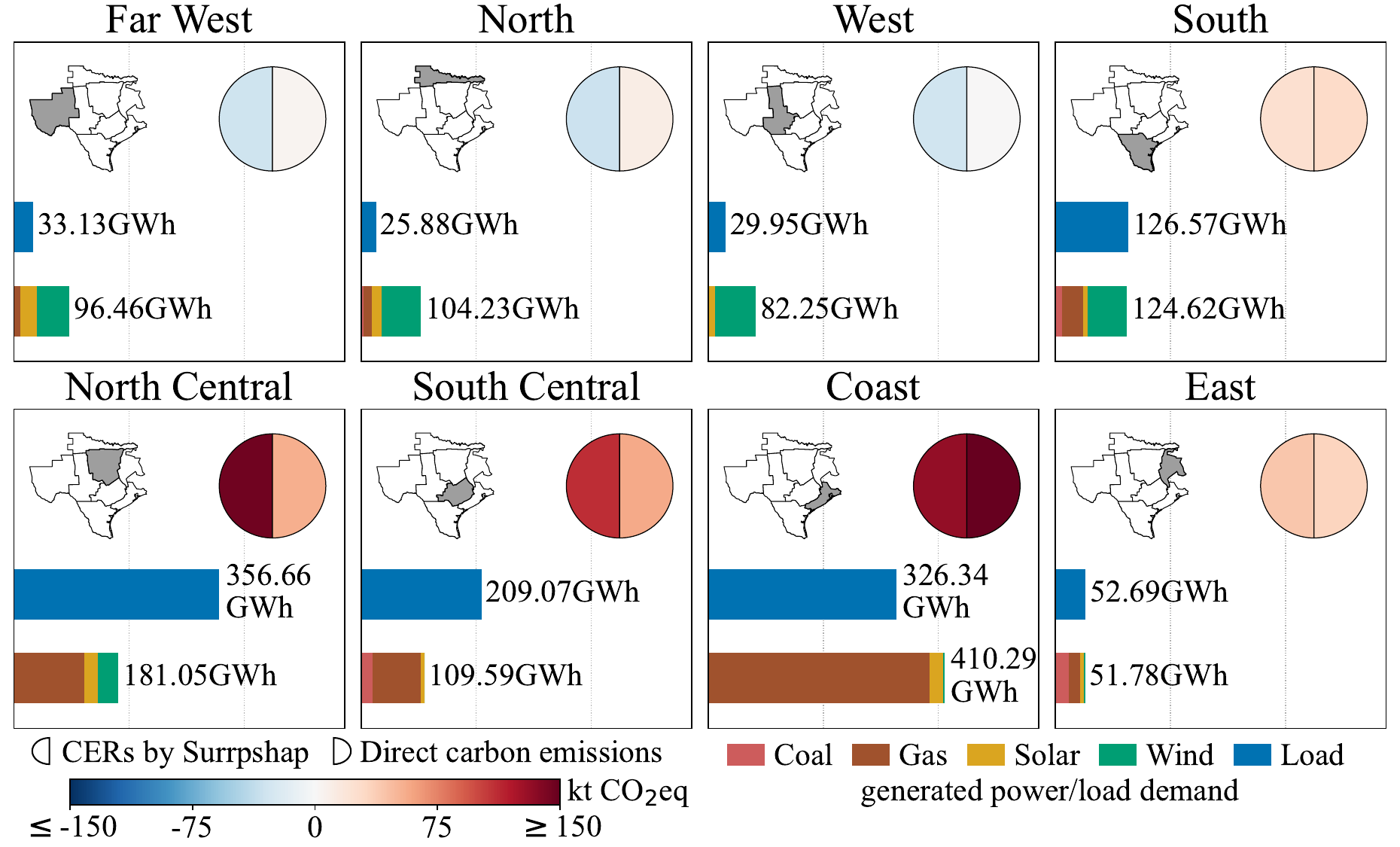}
\caption{Regional CER allocation across ERCOT weather zones. Circles show CERs versus direct emissions; bars indicate generation mix and demand.}
\label{ercot_8panel_style}
\end{figure}

Regional analysis in Fig. \ref{ercot_8panel_style} reveals geographic patterns (daily avereage over the year) across ERCOT's eight weather regions. Each region panel shows the pie chart of CER allocation (left) versus direct carbon emissions (right), with bar charts depicting daily generation by technology and load demand. Renewable-rich regions (Far West, North, West) achieve negative total CERs despite hosting thermal generation, as their renewable exports offset local emissions. Load centers (North Central and South Central) incur substantial positive CERs exceeding their direct emissions, reflecting their role in driving system-wide generation. The Coast region, despite high direct emissions from substantial thermal generation, shows moderated CER as approximately 20\% of its generation exports to other regions, demonstrating SurroShap's network awareness.

\subsection{Comparison with Alternative Allocation Methods}\label{Section4G}
Having demonstrated that SurroShap accurately approximates exact Shapley values, we use it as a fairness benchmark to evaluate three CER allocation methods in literature: carbon emission flow (CEF) \cite{kang2015carbon,7741580}, marginal carbon intensity (MCI) \cite{valenzuela2023dynamic}, and Aumann-Shapley (AS) \cite{chen2018method,xie2024real}. CEF and MCI assign all responsibility to loads (zero for generators), while AS enforces equal generation-demand splits. In contrast, SurroShap allocates directly to individual entities, potentially assigning over 100\% to demand-side entities while some generators receive negative allocations (see Section \ref{Section4F}).

\begin{figure}[!t]
\centering
\includegraphics[width=88mm]{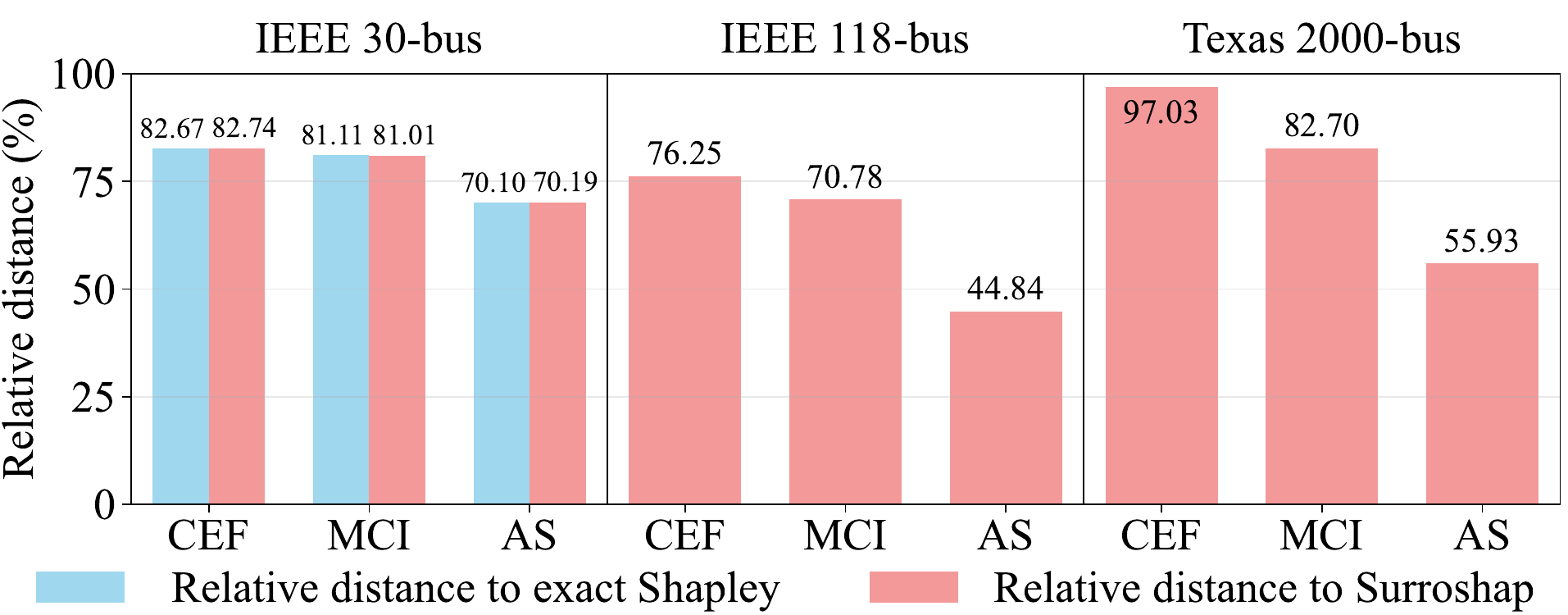}
\caption{Relative distance of alternative CER allocation methods from Shapley-based fairness benchmarks.}
\label{dif_method_error}
\end{figure}

Fig. \ref{dif_method_error} compares daily CER allocations (aggregated from hourly computations) across three systems, showing relative distances (L2 norm of allocation differences normalized by SurroShap's L2 norm). On IEEE 30-bus where exact Shapley is tractable, all alternatives show substantial deviations: CEF (82.67\%), MCI (81.11\%), and AS (70.10\%) from exact Shapley, with nearly identical distances when measured against SurroShap. For larger systems, deviations persist with AS showing 44.84\% (IEEE 118-bus) and 55.93\% (Texas 2000-bus) distance. These substantial gaps (45-97\%) demonstrate that existing methods are not capturing fairness properties, underscoring SurroShap's necessity for theoretically grounded fair allocation at scale.

\section{Conclusion}\label{Section5}
This work bridges the long-standing gap between the theoretical promise of Shapley value-based fair allocation and its practical deployment in large-scale power systems. By synergistically combining Kernel SHAP with deep learning acceleration, SurroShap transforms a computationally intractable problem into one solvable within operational timescales while maintaining provable error bounds. Multi-period analysis on the IEEE 30-bus system shows SurroShap's approximation error decreasing from 0.64\% to 0.17\% over multiple allocation rounds, remaining well below the theoretical bound of 0.37\%. The year-long Texas 2000-bus simulation reveals striking allocation patterns: loads bear 164.7\% of total emission responsibility while renewable generation offsets 63.9\%, quantifying how fair emission responsibilities propagate through large networks. Computational performance validates real-world applicability, with the 2000-bus system completing allocations in 3.17 minutes compared to months required by previous methods. Furthermore, existing CER allocation methods show substantial deviations of 45-97\% from Shapley-based fairness benchmarks, indicating the necessity of our approach. Future work may extend the SurroShap framework to other allocation problems in power systems and generalize the surrogate modeling approach to handle characteristic functions involving equilibrium computations or systems of equations beyond optimization problems.

\bibliography{References}

\vfill

\end{document}